\documentclass[12pt]{article}
\usepackage{latexsym,amsfonts,amsmath,amsthm,amssymb}
\usepackage{amsmath}
\usepackage{amssymb}
\usepackage{amscd}

\title{Quatum-like model of processing of information in the brain based on classical electromagnetic field}
\author{Andrei Khrennikov\\
\\ International Center for Mathematical Modelling
\\in Physics and Cognitive Sciences,\\
Linnaeus University, S-35195, Sweden}

\begin{document}
\maketitle

\abstract{We propose a model of quantum-like (QL)  processing of mental 
information. This model is based on quantum information theory. However, in contrast to models 
of ``quantum physical brain'' reducing mental activity (at least at the highest level) 
to quantum physical phenomena in the brain, our model matches well with the basic neuronal paradigm of the
cognitive science. QL information processing is based (surprisingly) on classical 
electromagnetic signals induced by joint activity of neurons. This novel approach to quantum information 
is based on representation of quantum mechanics as a version of classical signal theory which was recently 
elaborated by the author. The brain uses the QL representation (QLR) for working with abstract concepts;
concrete images are described by classical information theory. Two processes, classical and QL, are performed 
parallely. Moreover, information is actively transmitted from one representation to another. A QL concept given in our model
by a density operator can generate a variety of concrete images given by temporal realizations of the corresponding (Gaussian) random signal.
This signal has the covariance operator coinciding with the density operator encoding the abstract concept under consideration.
The presence of various temporal scales in the brain plays the crucial role in creation of QLR in the brain. Moreover,
in our model electromagnetic noise produced by neurons is a source of superstrong QL correlations between processes in different spatial 
domains in the brain; the binding problem is solved on the QL level, but with the aid of the classical background fluctuations.
}         

 \section{Introduction}
Last years the hypothesis that the brain processes information (at least partially) by using quantum-like (QL) representation
of probabilities by complex amplitudes was discussed from various viewpoint by numerous authors, e.g., \cite{Hameroff1}--\cite{KHR0}, see also \cite{Conte0}-- \cite{Conte}
for statistical studies confirming (at least preliminary)  this hypothesis.
In general such processing need not be based on the physical quantum brain (cf., e.g., Homeroff, Penrose, and Vitiello, see 
\cite{Hameroff1}--\cite{Vit})  --
quantum physical carriers of information. In our approach the brain
created the QL representation (QLR)
of information in Hilbert space. It uses quantum information rules in decision making. The existence
of such QLR was (at least preliminary) confirmed by experimental data from cognitive psychology and economics.\footnote{In \cite{Conte0}--
\cite{Conte} we performed experiments with recognition of ambiguous figures which demonstrated interference effects for 
incompatible recognition tasks in the form of 
violation of the law of total probability (LTP), see \cite{KHRIP}, \cite{KHRC}, \cite{KHR0} for details. 
(One can speak about non-Kolmogorovness of probabilistic data, i.e., impossibility to describe it by using 
the Kolmogorov probability model \cite{K}.) In \cite{Jerome1}--\cite{Jerome3}, \cite{KHR0} it was demonstrated that statistical data 
from well known experiments in cognitive economics, Shafir and Tversky \cite{ST}, \cite{TS}, on so called disjunction effect in decision making
also exhibit interference effect by violating LTP. We recall that the disjunction effect is related to violation of 
Savage's sure thing principle \cite{S} -- the basic principle of the modern economic theory, the principle of rationality of decision making.}
The violation of LTP in these experiments is an important sign of nonclassicality of data -- the interference 
effect. Moreover, recently so called  {\it constructive wave function approach} was developed:  data violating LTP  
can be represented by complex probability amplitudes by application of a special algorithm -- QLR-algorithm \cite{KHRC}, \cite{KHR0}.  
Recently there were also developed  QL models of decision making, see \cite{Jerome1}--\cite{Jerome3},  \cite{Khrennikov/QLBrain}, 
\cite{Bio2009}, \cite{AS}, .

 The next natural step is to try to find  possible physical realizations of QLR
in the brain. One of possibilities is to appeal to quantum physics of microprocesses in the brain --
the quantum brain, see, e.g., Homeroff and Penrose, see \cite{Hameroff1}--\cite{PEN2}. However, (surprisingly) it is possible to proceed even in the classical field framework, i.e.,
to create a classical wave model producing  QLR, see \cite{PCSFT1}--\cite{PCSFT4}.
Neurophysiologically our model is based on  {\it variety of time scales in the brain}, see e.g. \cite{Geissler}.
Each pair of  scales (fine -- the background fluctuations of electromagnetic field and  rough -- the mental image scale)
induces the QL representation. The background field plays the crucial role in creation
of ``superstrong QL correlations'' in the brain.

We propose a classical (!) wave model which reproduces
probabilistic effects of quantum information theory.
Why do we appeal to classical electromagnetic fields in the brain and not to quantum phenomena? In neurophysiological and
cognitive studies  we see numerous classical electromagnetic waves in the brain. Our conjecture is that these waves are
carriers of mental information which is processed in the framework of quantum information theory.

In the quantum community  there is a general opinion that quantum effects
can not be described by classical wave models (however, cf. Schr\"odinger; we also can mention works of H. F. Hofmann 
who demonstrated that one can use classical electromagnetic fields to describe a variety of ``purely cquantum effects'' \cite{Hofmann}.). Even those who agree that the classical and quantum  interferences  are similar emphasize the role of quantum entanglement and its  irreducibility to classical correlations (however, cf.
Einstein-Podolsky-Rosen). It is well known that entanglement is crucial in quantum information theory. Although some authors emphasize the role
of quantum parallelism in quantum computing, i.e., superposition and interference,
experts know well that without entanglement the quantum computer
is not able to beat the classical digital computer.

Recently  the author proposed
a classical wave model reproducing all probabilistic predictions of quantum mechanics, including correlations
of entangled systems, so called {\it prequantum classical statistical field theory} (PCSFT)\cite{PCSFT1, PCSFT3}
and see paper\cite{PCSFT4} for the recent model for composite systems.
It seems that, in spite of the mentioned common opinion, the classical wave description of quantum phenomena is still possible.

In this paper we apply PCSFT to model QL processing of information in the  brain on the basis of classical electromagnetic fields.
This model is based on the presence of various {\it time scales in the brain.}
Roughly speaking each pair of time scales, one of them is fine -- the background fluctuations of electromagnetic (classical) field in the brain, and another is rough -- the mental image scale,
can be used for creation of QLR in the brain. The background field (background oscillations in the brain) which is an important part of our model
plays the crucial role in the creation
of ``superstrong QL correlations'' in the brain, cf. Hofmann \cite{Hofmann}. These mental correlations are nonlocal due to
the background field. These correlations might provide a solution of  the {\it binding problem.}

Each such a pair of time scales, (fine, rough), induces QLR of information.
As a consequence of variety of time-scales in the brain, we get a variety of QL representations
serving for various mental functions. This QL model of brain's functioning was originated
in author's paper\cite{KHRCS}. The main improvement of the ``old  model'' is due to a new possibility
achieved recently by PCSFT: to represent the quantum correlations for entangled systems as the correlations
of the classical random
field, so to say prequantum field. This recent development also enlighted the role of the {\it background field}, vacuum fluctuations. We now
transfer this mathematical construction designed for quantum physics to the brain science.  Of course, it is a little bit naive model,
since we do not know the ``QL code'' used by the brain: the correspondence between images and probability distributions of
random electromagnetic fields in the brain.

We speculate that decision making through nonclassical LTP is based on a  {\it wave representation of information}
in the brain.   The brain is  full of  classical electromagnetic radiation. May be the brain was able to create QLR of information
via classical electromagnetic signals, cf.  K.-H. Fichtner, L. Fichtner, W. Freudenberg and M. Ohya \cite{Wolfgang}.
We also make a remark on the approach of G. Vitiello \cite{Vit}. It differs essentially from majority of  approahces 
to quantum physical brain. In contrast to others, Vitiello uses not quantum mechanics, but quantum field theory, In some sense 
his model is closer to PCSFT. He also operate with macroscopic electromagnetic signals, but for him these signals are described by 
the quantum field model.
 
It is well known that classical waves produce superposition and, hence,  violate LTP. However,
quantum information processing is based not only on
superposition, but also on  ENTANGLEMENT. It is the source of  superstrong nonlocal correlations.
Correlations are really superstrong -- violation of Bell's
inequality.   Can  entanglement be produced by classical signals? Can quantum information processing be reproduced
by using classical waves? Surprisingly, the answer is positive.

The crucial element of our classical wave model of brain's functioning is the presence of the random background field
(in physics fluctuations of vacuum, in the cognitive model --
background fluctuations of the brain). Such a random background increases essentially correlations between different mental functions,
generates nonlocal presentation of information. As was already remarked,
we might couple these nonlocal representation of information to the binding problem:

{\it ``How the unity of conscious perception is brought about
by the distributed activities of the central nervous system.''}

\section{Why may brain use the quantum-like representation of information
based on classical electromagnetic waves?}

As we emphasized, the deep neurophysiological studies demonstrated that the brain definitely processes the information
by using classical electromagnetic signals. We would like to apply the results of these studies and propose a classical-signal
model of the brain functioning. However, we do not plan just to explore the standard classical
signal theory.  We speculate that information processing in the brain should be described by the mathematical formalism
of quantum mechanics and that classical electromagnetic waves are used by the brain to create QLR.

\medskip

``Why was the brain not satisfied with the classical signal processing?

\medskip

What are advantages of the QL processing of information
(even with classical field)?''

\subsection{Incomplete processing of information}

If we speculate  that in physics quantum probabilistic behavior can be
expressed through classical random waves, we definitely reject Bohr's thesis on completeness
of QM. (This Einsteinian attitude is characteristic only for this chapter, in previous chapters
we were able to proceed even with the orthodox Copenhagen interpretation.)
Here we have to assume that the QM formalism provides only an approximate description of processes in the
micro world. In such an approach the main difference of quantum processing of information from classical
is that the first one provides a possibility to ignore consistently a part of information, to make
a consistent cutoff of information (described by the mathematical formalism of QM).

Operation through incomplete information processing is very profitable for cognitive systems. An important
part of cognition is the extraction of a part of information from huge information flows coming
to the brain. By operating in the QL-framework, for example, on the basis of the wave representation,
the brain gets a possibility to work harmonically with incomplete information. This is one of sources
of creation of the QL- processing in the brain.

By ignoring a part of information the brain is able to create abstract mental images, ideas, concepts,
categories. Another advantage is incredible increasing of the speed of computations. Here we speak about
computations based on classical electromagnetic signals, but performed on the basis of quantum formalism.

\subsection{Background noise: How can the worst enemy become the best friend?}

By PCSFT the standard QM formalism provides a possibility to extract signals (in fact, their averages)  from the
noisy background, see section \ref{STR}.
QM  can be interpreted as a kind of renormalization theory which is applicable to signals with irreducible noise. In quantum physics this
 is the noise of vacuum fluctuations, the background field. It seems that this noise is a fundamental feature of space \cite{H1},
\cite{H2}; one cannot hope to isolate
a signal from these fluctuations; the only possibility is to take them into account consistently. We now project this situation to
brain's functioning. The brain is a complex electric system; its functioning has definitely to induce noise; more complex brains
are more noisy; higher brain's activity is also noisier. Of course, in the process of evolution cognitive systems might try
reduce the impact of the background noise in the brain. However, it is clear that it would be a very complicated task: it seems
impossible to isolate signals in the brain from the background fluctuations (induced by a huge number of neurons).

We speculate that the brain chosen another way to evolve, namely, to elaborate a procedure to extract mental images
from signals on the noisy background, i.e., proceeding with the noise, but, finally, having a procedure (consistent for different signals)
for extracting images. By PCSFT processing of information on the basis of the QLR provides such a possibility.

Moreover, in physics (by PCSFT) the background field is a source of superstong nonlocal correlations between entangled systems -- ``entangled classical
waves'' in the PCSFT-framework. By projecting this situation to the brain functioning we see that the brain can get the
great advantage from the presence of the background fluctuations. They produce entanglement between processes in different
(including spatially separated) domains of the brain, between different mental functions, see section \ref{ENTPR} for further discussion.

\subsection{Joint processing of an abstract image and its concrete realizations}
\label{PIPI}

By PCSFT a random electromagnetic field is represented in the QL-way by its covariance operator (=``density operator'').
In the cognitive model the QL-representation corresponds to abstract mental images. Thus they are given by covariance
operators. To process such QL-images the brain is not interested in the complex structure of random fluctuations of
classical signals; the brain operates
with images encoded by operators -- covariance operators (matrices) of classical random signals. However, in this process the brain might need to proceed
from the abstract image to its concrete realization. For example, the brain can operate with abstract notions,
e.g., house and tree, but it can switch to the concrete house and the concrete tree. In processing of the first type (the QL-processing) the
brain operates with operators $D_{\rm{house}}$ and $D_{\rm{tree}}$ and in processing   of the second type (the classical processing)
the brain has to switch to the classical signals encoding this house and this tree.

We remark that two different realizations of the
random signal with the fixed covariance operator can differ essentially, so they really can encode two different houses.
Take a coin. Consider a series of its tossing, e.g., a few thousands: $x=(x_1,x_2,...., x_n),$ where $x_j=0,1$ are labels
of coin's sides. After this tossing was finished, start a new series: $y=(y_1,y_2,..., y_n).$ Although both sequences are samples
of the same random law, they can differ essentially bit wise. Thus we can encode ``house'' (as the abstract notion) by the probability law of this coin, but two concrete houses are encoded by sequences $x$ and $y.$

If the brain works in the QL (abstract) regime,
it recognizes only encoding of the corresponding probability laws;
if it works in the classical regime it has to use essentially more processing resources, since
it operates not with the codes of probabilistic laws, but with the real data streams.

In the QL-model induced by PCSFT we consider only {\it Gaussian random fields} with zero averages (symmetric oscilations). Such random fields are uniquely determined by covariance operators. The former remark on encoding by using the coin and its tossings can be modified in the following way.
We have a Gaussian random generator producing vector data, each vector (realization) has dimension  $m.$ Starting with some input
given by the vector $x_0$ the generator produces the stream of vectors $x$ encoding the concrete house. Starting with another
input $x_0^\prime$ the generator produces another stream of random data $y$  encoding another concrete house. The concept of house
is represented by the covariance operator $D =D_{\rm{house}}$ of this random generator, the density operator in the QL-formalism.

\section{Prequantum classical statistical field theory: noncomposite systems}
\label{PCSFT}

 Quantum mechanics (QM) is a statistical theory. It
cannot tell us anything about an individual quantum system, e.g.,
electron or photon. It predicts only  probabilities for results of
measurements for ensembles of quantum systems. Classical
statistical mechanics (CSM) does the same. Why are QM and CSM
based on different probability models?

In CSM averages are given by integrals with respect to probability
measures and in QM by traces. In CSM we have:
\begin{equation}
\label{AV0}
\langle f\rangle_\mu= \int_M f(\phi) d \mu(\phi),
\end{equation}
where $M$ is the state space.  In probabilistic terms: there is given a
random vector $\phi(\omega)$ taking values in $M.$ Then $\langle
f\rangle_\phi = E f(\phi(\omega))= \langle f\rangle_\mu.$
In QM the average is given by the operator trace-formula:
\begin{equation}
\label{AV1}
\langle \widehat{A} \rangle_\rho= \rm{Tr} \rho
\widehat{A}.
\end{equation}
This formal mathematical difference induces the prejudance on fundamental difference between classical and quantum worlds.
Our aim is to show that, in spite of the common opinion, quantum averages can be easily represented
as classical averages and, moreover, even correlations between entangled systems can be expressed as classical correlations
(with respect to fluctuations of classical random fields).

\subsection{Einstein's dreams:}
Albert Einstein did not believe in irreducible
randomness, completeness of QM. He dreamed of a better, so to say ``prequantum'', mode l\cite{EI}:

1). {\bf Dream 1.} A mathematical model reducing quantum
randomness to classical.

2). {\bf Dream 2.} Renaissance of causal description.

3). {\bf  Dream 3.} Instead of particles, classical
fields will provide the complete description of reality -- reality
of fields \cite{EI}:

``{\small But the division into matter and field is, after the recognition
of the equivalence of mass and energy, something artificial and
not clearly defined. Could we not reject the concept of matter and
build a pure field physics? What impresses our senses as matter is
really a great concentration of energy into a comparatively small
space. We could regard matter as the regions in space where the
field is extremely strong. In this way a new philosophical
background could be created."}

\medskip

The real trouble of the prequantum wave model (in the spirit of early Schr\"odinger)
are not various NO-GO theorems (e.g., the Bell inequality\cite{KHRC, KHR1}), but the problem which was recognized
already by Schr\"odinger. In fact, he gave up with his wave quantum mechanics, because of this problem:
{\it  A composite quantum system cannot be described by waves on physical space!}
Two electrons are described by the wave function on ${\bf R}^6$ and not by two wave on ${\bf R}^3.$

Einstein also recognized this problem \cite{EI}: ``{\small For one elementary particle, electron or photon, we have
probability waves in a three-dimensional continuum, characterizing
the statistical behavior of the system if the experiments are
often repeated. But what about the case of not one but two
interacting particles, for instance, two electrons, electron and
photon, or electron and nucleus? We cannot treat them separately
and describe each of them through a probability wave in three
dimensions...}''

\subsection{Quantum system = classical random field}

Einstein's Dreams 1 and 3 came true in PCSFT (but not Dream 2!) -- a version of CSM in which
fields play the  role  of particles.\footnote{It seems surprising that, although Dream 1 came true, Dream 2 cannot.
The situation differs essentially from CSM where dynamics of probability distribution given by
the Liouville equation can be reduced to the deterministic Hamiltonian dynamics. The main difference is due to the
presence of the background fluctuations -- irreducible noise.}
In particular, composite systems can be described by vector random fields, i.e.,
by the Cartesian product of state spaces of subsystems and not the tensor product.
The basic postulate of PCSFT can be formulated in the following way:

{\it A quantum particle is the symbolic representation of
a ``prequantum'' classical field  fluctuating on the time scale which is essentially finer than the time scale of measurements.}

The prequantum state space $M= L_2({\bf
R}^3),$ states are fields $\phi: {\bf R}^3 \to {\bf R};$
``electronic filed'', ``neutronic field'', ``photonic field'' -
classical electromagnetic field.
An ensembles of ``quantum particles'' is represented by an
ensemble of classical fields, probability measure $\mu$ on $M=
L_2({\bf R}^3),$ or random field $\phi(x, \omega)$ taking values
in $M= L_2({\bf R}^3).$ For each fixed value of the random parameter $\omega= \omega_0,$
$x \to \phi(x, \omega_0)$ is a classical field on physical space.

\subsection{Density operator = covariance operator}

Each measure (or random field) has the covariance operator, say
$D.$ It describes correlations between various degrees of freedom.

The map $\rho \mapsto D= \rho$ is one-to-one between density operators
and  the covariance operators of the corresponding prequantum
random fields -- in the case of noncomposite quantum systems. In the case
of composite systems this correspondence is really tricky.

Thus each quantum state (an element of the QM formalism) is represented by the classical random
field in PCSFT. The covariance operator of this field is determined by the density operator. We also postulate that the
prequantum random field has {\it zero mean value.} These two conditions determine uniquely Gaussian random fields.
We restrict our model to such fields. Thus by PCSFT quantum systems are Gaussian random fields.

Finally, we remind that the covariance operator $D$ of a random field $\phi$ is defined by its bilinear form
($u,v \in H):$
\begin{equation}
\label {COV}
\langle D u, v \rangle= E \rangle \phi, u\langle \langle v,  \phi\rangle=
E \left( \int_O \phi(x, \omega) \overline{u(x)} dx\right)
\left( \int_O  v(x) \overline{\phi(x, \omega)} dx\right)
\end{equation}
or by using the probability distribution $\mu$ of the random field:
\begin{equation}
\label {COV1}
\langle D u, v \rangle=
\int_H \left( \int_O \phi(x) \overline{u(x)} dx\right)
\left( \int_O  v(x) \overline{\phi(x)} dx\right) d \mu(\phi).
\end{equation}

\subsection{Quantum observable = quadratic form}

 The map $\widehat{A} \to f_A(\phi)= (\widehat{A} \phi, \phi)$ establishes one-to-one correspondence
between quantum observables (self-adjoint operators) and classical physical variables (quadratic functionals of
the prequantum field).

It is easy to prove that following equality holds:
\begin{equation}
\label{AV3}
E f_{A}(\phi_ (\omega))= \int_M f_A(\phi) d \mu(\phi)= \rm{Tr}
\rho \widehat{A}.
\end{equation}
In particular, for a pure quantum state $\psi,$ consider the Gaussian
measure with zero mean value and the covariance operator $\rho=
\psi\otimes\psi$ (the orthogonal projector on the vector $\psi),$ then
$$
\int_M f_A(\phi) d \mu(\phi)= ( \widehat{A} \psi, \psi).
$$
This mathematical formula coupling integral  of a quadratic form and the corresponding trace is well known
in measure theory. Our main contribution is coupling of this mathematical formula with quantum physics.

This is the end of the story for quantum noncomposite systems, e.g., a single electron or photon\cite{PCSFT1, PCSFT3}.

\subsection{Beyond QM}

In fact, PCSFT not only reproduces quantum averages, but it also
provides a possibility to go beyond QM. Suppose that not all prequantum physical variables are given by
QUADRATIC forms, consider more general model, all smooth
functionals $f(\phi)$ of classical fields. We only have the
illusion of representation of all quantum observables by
self-adjoint operators.

The map
\begin{equation}
\label{AV3Z0}
f \mapsto \widehat{A}= f^{\prime\prime}(0)/2
\end{equation}
  projects smooth functionals of the prequantum field (physical variables in PCSFT ) on self-adjoint operators (quantum observables). Then quantum and classical (prequantum) averages do not coincide precisely, but only approximately:
\begin{equation}
\label{AV3Z}
\int_M f_A(\phi) d \mu(\phi)= \rm{Tr} \rho \widehat{A} + O(\tau/T),
\end{equation}
where $T$ is the time scale of measurements and $\tau$ the time scale
of fluctuations of prequantum field. The main problem is that PCSFT does not provide a quantative estimate of the time scale of fluctuations of the
prequantum field. If this scale is too fine, e.g., the Planck scale, then QM is ``too good approximation of PCSFT'', i.e., it would be really impossible to distinguish them experimentally. However, even a possibility to represent QM as the classical wave mechanics can have important
theoretical and practical applications. And in the present paper we shall use the mathematical formalism of PCSFT to model
brain's functioning.  Although even in this case the choice of the scale of fluctuations is a complicated problem, we know
that it is not extremely fine; so the model can be experimentally verified (in contrast to Roger Penrose we are not looking for
cognition at the Planck scale!).

\section{Cognitive model: Two regimes of brain's functioning}

We now turn to considerations of section \ref{PIPI} and proceed on
the basis of  the short presentation of PCSFT given in section
\ref{PCSFT}. At the moment we consider one fixed mental function
of the brain, say $F,$ which is physically concentrated in some
spatial domain $O \subset {\bf R}^3$ of the brain. We shall come
to the model of QL cooperation of a few mental functions after the
presentation of PCSFT for composite systems, section \ref{CS}
(``entanglement of mental functions'').

\subsubsection{Classical regime}
\label{CR}

By getting an input $\phi_0$ (from environment or another mental
function) the mental function $F$  produces a random signal
$\phi(x,\omega)$ -- a classical electromagnetic field resulting
from neuronal activity.\footnote{Thus we consider the ensemble of
neurons which are located in $O.$ By our model the brain does not
so much interested in the ``private life'' of individual neurons,
i.e., the frequency of spikes and so on, cf. \cite{DA}. It is only
interested in the electromagnetic field induced by activity of
these neurons.}
 It is a random signal depending on the chance parameter $\omega.$ For each
$\omega_0,$ this electromagnetic field, $x \to \phi(x,\omega_0),$
is distributed on the domain $O.$

We use the complex
representation for the electromagnetic field, the
Riemann-Silberstein representation:
$$
\phi(x)= E(x) + i B(x),
$$
where $E(x)= (E_1(x), E_2(x), E_3(x))$ and $B(x)=(B_1(x), B_2(x), B_3(x))$
are the electric and magnetic components, respectively.

In our model {\it each  concrete mental image
is associated with a random signal.} Its mental features, {\it ``qualia''}, are given
by functionals of this signal. In the simplest case these are
quadratic forms of the signal.

The main problem is to create the classical signal code, i.e., to
establish correspondence between random signals and mental images
as well as between field-functionals and qualia of images. We
speculate that at least some field-functionals represent {\it
emotions} related to the mental image (which is represented by the
classical electromagnetic signal). Consider a number of emotions,
say ${\cal E}_1,..., {\cal E}_k,$ related to some image, say
$MI_\phi$ (associated with the signal $\phi).$ Then the mental
function $F$ physically operates with the corresponding
field-functionals; in the simplest case these are quadratic
functionals and they can be represented by integral kernels:
\begin{equation}
\label{IK} f_{{\cal E}_j}(\phi)= \int_{O\times O} K_j(x,y) \phi(x)
\overline{\phi(y)} dx dy.
\end{equation}
However, in the classical regime {\it nonquadratic functionals} are also
in the use; e.g.,
$$
f(\phi)= f_{{\cal E}}(\phi)+ \int_{O\times O \times O} K(x,y,z) \phi(x)  \phi(y)
\overline{\phi(z)} dx dy dz,
$$
where $f_{{\cal E}}$ is the functional  of the form (\ref{IK}).

\medskip

{\bf Remark 1.}  (Spatial distribution of qualia) By considering integral functionals of the classical electromagnetic field
we suppose that the $F$-function performs integration of a signal over its  domain
of spatial concentration. Thus we consider only spatially concentrated mental functions.
If a mental function $F$ is concentrated in a domain $O=\cup_k O_k,$ where $O_k$ are located
in the brain  far away from each other, then we represent $F$ as collection of ``elementary
mental functions'' $F_k$ concentrated in domains $O_k.$ Some qualia of $F$ are associated with
elementary functions $F_k.$ However, there are also exist global qualia which are obtained by summation of local
ones (so integration on each $O_k$ and then collection and summation in a special center).

\medskip

Since signals are random, field functionals  are fluctuating quantities -- random variables:
$\xi_\phi(\omega)= f(\phi(\omega)).$ It is
clear that the brain cannot operate with such unstable mental
entities. Thus it has to produce averages of emotions and operate
with them. It will be especially clear in the time-representation
of random signals, see section \ref{TRTR}. Let $\mu$ be the
probability distribution of a random signal $\phi.$
Then emotions are quantified by averages:
\begin{equation}
\label{EOQ} \langle f_{{\cal E}_j} \rangle = \int_H  f_{{\cal
E}_j}(\phi) d \mu(\phi),
\end{equation}
where $H=L_2(O).$ In our model not only emotions, but all qualia
are quantified by averages.

It is clear that quantification of each qualia consumes brain's
resources. Therefore only a special class of qualia (in
particular, emotions) is associated with each mental image. How
does the mental function $F$ select them is the open question. The
crucial point is that in principle any two emotions ${\cal E}_1$
and ${\cal E}_2$  or other qualia can be associated with the image
$MI_\phi$ and  quantified.  This {\it total compatibility of
emotions and qualia in general} may induce some problems. For
example, it is not always profitable for survival to combine some
emotions. We shall see that the situation is totally different in
 QL processing of information.

Functionals of classical electromagnetic signals represents not
only emotions, but even other qualia of the image $MI_\phi.$
For example, in PCSFT we have the energy variable (representing the
intensity of a signal):
\begin{equation}
\label{EO} f_I(\phi)= \int_O \vert
\phi(x)\vert^2 dx =\int_O (E^2(x)+B^2(x)) dx.
\end{equation}
We relate this functional to the {\it intensity of feeling} of the
image $MI_\phi.$ This intensity is quantified as
\begin{equation}
\label{EOQ} \langle f_I \rangle = \int_H f_I(\phi) d \mu(\phi)=
\int_H\left(\int_O \vert \phi(x)\vert^2 dx\right)d \mu(\phi),
\end{equation}
$$
 =\int_H\left(\int_O (E^2(x)+B^2(x)) dx\right)d \mu(\phi).
$$
In QM the position observable is given by the multiplication
operator
$$
\widehat{x} \phi(x) = x \phi(x)
$$
and in PCSFT it is represented by the field functional:
\begin{equation} \label{EO1} f_x(\phi)= (\widehat{x} \phi, \phi)=
\int_O x \vert \phi(x)\vert^2 dx =\int_O x (E^2(x)+B^2(x)) dx.
\end{equation}
What quale can be coupled to this functional?

Already here, on the level of classical mental processing,
mathematics lead us to the notion of {\it conjugate qualia
of a mental image.} For example, consider the momentum functional in
PCSFT:
\begin{equation}
\label{EO1} f_p(\phi)= (\widehat{p} \phi, \phi)= \int_{{\bf R}^3}
p \vert \tilde{\phi}(p)\vert^2 dp,
\end{equation}
where $\tilde{\phi}(p)$ is the Fourier transform of the signal
$\phi(x).$ What is a cognitive interpretation of conjugation between qualia given by functionals
$f_x$ and $f_p?$

It seems that in the classical regime the brain can process
conjugate qualia simultaneously. In the case of
``position and momentum'' functionals it is  simultaneous
processing in the spatial and frequency representations.

\medskip

{\bf Classical mental coding:} How does the brain associate the
mental image $MI_\phi$ with a classical signal $\phi?$ In our
model it is done  through calculation of its covariance operator
$D=D(\phi).$ This association, mental image -- covariance
operator, is especially natural in the time representation of
random signals, see section \ref{TRTR}.

\subsubsection{Quantum-like regime}

We are now interested in the QLR (quantum-like representation) of
information. In QLR the brain operates with density operators
which represent not only concrete mental images (coming from the
classical regime of mental processing), but also abstract concepts
(of different levels of abstraction, see section \ref{QLQL} for
details) which do not correspond to classically produced images.

In QLR the brain's state space
is space of density operators
${\cal D}(H), H=L_2(O).$ In principle each density operator can be
used as a QL state of the brain. However, it is natural to assume
that each mental function $F$ operates in its own subspace ${\cal
D}_F(H)$ of ${\cal D}(H).$

In the standard QM a system has not only the state, but also
``properties'' or (depending on interpretation) there are defined
observables on this system (in this state) -- e.g., the energy
observable, the coordinate observable and so on.  By QM they are
represented by self-adjoint operators.

To simplify mathematics we shall consider only bounded (continuous) operators; denote the space of
all bounded self-adjoint operators by the symbol ${\cal L}_s(H).$
For a given quantum state $\rho \in {\cal D}(H)$ and observable $\widehat{A}\in {\cal L}_s(H),$ the QM formalism
gives the average of this observable in this state, see (\ref{AV1}).

We encode qualia of a QL cognitive image $MI_\rho$ (which is encoded by a density operator
$\rho)$ by self-adjoint operators; thus qualia of a mental image (in QLR of information)
are described by the space ${\cal L}_s(H).$ They are quantified by their averages, via (\ref{AV1}).

Opposite to classical processing, in QLR the brain cannot quantify
all qualia simultaneously. There exist {\it
incompatible qualia}; in particular, incompatible
emotions. The QL brain can select different representations of the
mental image $MI_\rho$  and different collections of compatible
qualia of the image. The QL brain escapes the
simultaneous use of e.g. some emotions (``incompatible
emotions''). In this way QLR-processing differs essentially from
classical processing. We can speculate that in the process of
evolution the brain created (on the basis of experience)
commutative algebras corresponding to compatible qualia.

\subsubsection{Coupling between classical and quantum-like representations}
\label{LKP}
The crucial point of QLR of information is that this ``operator-thinking'' is
naturally coupled with processing  of classical electromagnetic signals.  On the level of mental images we have:

\medskip

aMI). From the classical regime to QL: a classical signal $\phi$
induces the mental image $MI_\phi$ given by its covariance
operator $D=D(\phi)$ and it is transferred to QLR through
normalization by the trace: $D\to \rho= D/\rm{Tr} D.$ Thus there is
a map from classical mental images to QL mental images: $MI_\phi
\to MI_\rho.$

bMI). From QL  to classical: a QL mental image $MI_\rho$ can be
represented by a classical (Gaussian) signal $\phi=\phi_\rho$ with
the covariance operator $\rho,$ i.e., by the image $MI_\phi.$

\medskip

On the level of qualia we have:

\medskip

aQU). From classical regime to QL: each functional of classical
field, $f(\phi),$ is represented by its second derivative --
self-adjoint operator, see (\ref{AV3Z0}).

bQU). From QL  to classical: each quantum quale (given by a self-adjoint operator) is represented
by its quadratic form.

\medskip

Since aQU is not one-to-one, i.e., since a huge
class of different classical qualia given by various
functionals of the electromagnetic field (all functionals with the same second
derivative) is mapped into the same
self-adjoint operator, QLR makes the mental picture less rich than it
was in the classical representation. The same
can be said about aMI. Various classical signals, concrete mental
images, can have the same covariance operator (we do not claim
that input signals are obligatory Gaussian, so there is no
one-to-one correspondence), cf. with a general discussion on
mental prespace \cite{?}.

Take a classical mental image $MI_\phi.$ It is represented by the covariance operator $D=D(\phi).$ Of course, it can be mapped to a QL image $MI_\rho,$ see aIM. Values of all QL qualia
can be obtained by scaling from values of corresponding classical qualia, since, for any operator  $\widehat{A} \in {\cal L}_s(H),$
$$
\langle \widehat{A} \rangle_\rho= \rm{Tr} \rho  \widehat{A} = \frac{1}{\rm{Tr} D} \rm{Tr} D  \widehat{A}
$$
$$
= \frac{1}{\rm{Tr} D} \int_H f_A(\phi) d\mu(\phi) = \frac{1}{\rm{Tr} D}  \langle f_A\rangle_\phi,
$$
where   $\mu$ is the probability distribution of the signal. We remark that
$$
\rm{Tr} D = \int_H \left( \int_O \vert \phi(x)\vert^2 dx \right) d \mu(\phi)= \langle f_I \rangle
$$
is the intensity of the signal or in the cognitive model the
intensity of feeling of the mental image $MI_\phi.$ Thus QL qualia
are normalized by the intensity of feeling:
 $$
\langle \widehat{A} \rangle_\rho= \frac{\int_H f_A(\phi) d\mu(\phi)}{\int_H \left( \int_O \vert \phi(x)\vert^2 dx \right) d \mu(\phi)}.
$$

QL processing of all mental images is performed on the same level
of intensity of feeling, so it is {\it ``calm thinking''.}

We remark once again that some classical qualia do not have quantum counterpart.

\section{Classical regime: Time representation}
\label{TRTR}

As usual in signal theory, we can switch from the ensemble
representation for averages to the time representation (under the
standard assumption of {\it ergodicity}). Thus, instead of a
random field $\phi(x,\omega),$ which  is distributed with some
probability distribution $d \mu(\phi)$ on $H,$ we consider a  time
dependent signal
$$\phi(s)\equiv \phi(s,x),$$
 where $x \in O, s\in [0, +\infty).$
Then, for each functional $f(\phi)$
such that $\int_H \vert f(\phi) \vert d\mu(\phi) < \infty,$ we have (by ergodicity):
\begin{equation}
\label{EOQ!}
\langle f \rangle_\mu\equiv \int_H  f(\phi)  d\mu(\phi)= \lim_{T \to \infty} \frac{1}{T} \int_0^T f(\phi(s)) ds \equiv \langle f \rangle_\phi.
\end{equation}
Consider two time scales: $\tau$ is a fine scale and $T>>\tau$ is a rough time scale. In QM the latter is the scale of measurements and
$\tau$ is the scale of fluctuations of the prequantum field.\footnote{PCSFT does not predict the magnitude
of the scale of prequantum field fluctuations. One may speculate (motivated in particular by cosmology and
string theory), cf. G. `t Hooft \cite{H1}, \cite{H2} that it has to be the Planck scale $\tau_{P}\approx 10^{-44}$ s.
If it were really the case, then prequantum fluctuations have only a 
theoretical value: they will be never approached experimentally. However, in \cite{IJTP}
we discussed a possibility that the prequantum  scale may have a larger magnitude and, hence, fluctuations will be soon or later
approached experimentally.}
In cognitive science we use the following interpretation of time scales:
$T$ is the scale of the QLR and $\tau$ is the scale of the real physical processing of the electromagnetic signal in the brain. Thus
\begin{equation}
\label{EOQ!!}
\langle f \rangle_\phi \approx \frac{1}{T} \int_0^T f(\phi(s)) ds,
\end{equation}
where $s$ denotes the time variable at the $\tau$-scale. We call
the $T$-scale the {\it mental time scale;} we can also speak about
{\it psychological time.} The $T$-scale is the scale of creation
of mental images by the brain. The $\tau$-scale is the physical
processing scale or {\it premental time scale.} Neurophysiological
theoretical and experimental studies provide the following
estimate of realtive magnitudes of these time scales. If we select
$\tau= 1$ mls., then $T\approx 80$ mls.

For each signal $\phi(s,x), x \in O,$  the brain can find its qualia, e.g., the strength of
feeling of this image:
\begin{equation}
\label{EOQ!!}
\langle f \rangle_\mu \approx \frac{1}{T} \int_0^T \left(\int_O(E^2(s,x) +B^2(s,x)) dx\right) ds.
\end{equation}
In particular, emotions (special qualia) are given by such functionals, e.g., $f_{\rm{anger}}, f_{\rm{sadness}},...$
Our formal mathematical model cannot provide the form of concrete emotion-functionals. We hope that in future
it will be described as the result of neurophysiological and cognitive studies.

In the time representation the covariance operator (its bilinear form) of the signal $\phi(s,x)$ is given by
$$
(D u,v) =\lim_{T\to \infty} \frac{1}{T} \int_0^T \left(\int_O u(x) \overline{\phi(s,x)} d x \int_O  \phi(s,x) \overline{v(x)} dx\right) ds,
$$
\begin{equation}
\label{EOQ!!!}
\approx \frac{1}{T} \int_0^T \left(\int_O u(x) \overline{\phi(s,x)} d x \int_O  \phi(s,x) \overline{v(x)} dx\right) ds,
\end{equation}
where $u(x), v (x)$ are two ``test signals'', $u, v \in L_2(O).$

\section{Classical Signal Processing of Mental Images}
\label{CL}

This section contains a detailed presentation of the classical
processing of information, see section \ref{CR}. We proceed in the
time representation of random signals.

 CSP1. {\bf Electromagnetic field basis of mental
images.} Inputs from external and internal worlds induce
electromagnetic signals in the brain.

Each signal has a variety of qualia; in particular, emotions
associated with the signal $\phi(s,x).$ Qualia are realized by
various functionals, $\phi \mapsto f(\phi),$ of the signal. They
are quantified by averages of these functionals, $f \mapsto
\langle f \rangle_\phi,$ see (\ref{EOQ!}) and (\ref{EOQ!!}). In
principle all possible qualia (e.g., emotions) can be jointly
associated with  $\phi(s,x).$

The physical dynamics of a signal is in general {\it nonlinear} and very complicated; it depends essentially
on context of the signal processing:
\begin{equation}
\label{PRSIG} \phi(s,x)= \phi(s_0,x) + \int_{s_0}^s d \alpha
\left(\int_O K(x,y,\alpha; \phi(\alpha,y); \phi(s_0,y)) dy
\right),
\end{equation}
where the kernel $K$ depends on the spatial variables $x,y \in O$ (the distribution of the signal on the brain), on
the time variable $\alpha,$ the previous dynamics of the signal $\phi(\alpha,y), \alpha \in [s_0, s),$ and the
signal at $s=s_0,$ the input. We remark that the dynamics $\phi(s,x)$ depends on input not only additively, i.e.,
as the initial state which then will evolve in accordance with some integral equation, but even the kernel of the equation
depends nontrivially on the input. Thus the dynamics are different for different inputs.
\medskip

CSP2. {\bf Calculation of correlations; creation of mental
images.} For each signal $\phi(s,x),$ the brain calculates the
corresponding covariance operator $D \equiv D(\phi),$ see
(\ref{EOQ!!!}). The completion of this process, i.e., calculation
of $D,$ is associated with creation of the {\it mental image}
$MI_\phi$ induced by the signal $\phi.$ Thus on the cognitive
level the brain is not interested in the dynamics of the physical
signal (\ref{PRSIG}). It is only interested in the dynamics of the
covariance operator
\begin{equation}
\label{VBN}
t \mapsto D(t).
\end{equation}
We remark that dynamics (\ref{PRSIG}) and (\ref{VBN}) have
different time scales; the first one is performed on the physical
time scale and the second on the mental time scale. Thus it is
very important that the ``physical brain'' and the ``cognitive
brain'' work on two different time scales: the scale of physical
signal -- $\tau,$ and the scale of QLR -- $T.$ The interval of
time $T >> \tau,$ so its size justifies the ergodic interplay
between ensemble and time representations of random signals.

\medskip

CSP3. {\bf Memory of correlations.} The density operator $D$ is
recorded in the brain. The PCSFT-basis of the model in combination
with the ergodic argument make very attractive the following model
of memory:

The operator $D$ determines uniquely the Gaussian probability distribution $\mu_{D}$ (with zero mean value).
The brain records this probability distribution.
How can it do this?

We speculate that, to encode $\mu_{D},$
the brain uses the statistical distribution by assigning statistical weights to elements of some
ensemble $\Omega.$

What are elements of $\Omega?$ They might be neurons or even distributions of chemical components in the brain.

We emphasize once again that such a model of memory for
probabilistic laws is based on the ergodicity of processes in the
brain: from a signal $\phi(s)\equiv \phi(s,x)$ (the time
representation) to its covariance and from the covariance to the
probability distribution on an ensemble.\footnote{The choice of a
Gaussian probability law can be debated. But at least
mathematically it works well, because of the one-to-one
correspondence between covariance matrices and Gaussian
probability distributions (with zero mean values).}

\medskip

CSP4. {\bf Recollection of images.} Recollection is the process of
activation of a special mental image.  We keep to the model of
statistical (ensemble) representation of the probability
distribution encoding the image, see CSP3. We obtain the following
procedure of recollection:

Suppose that a mental image $MI_\phi$ was recorded in the memory,
$\phi \mapsto D=D(\phi)\mapsto \mu= \mu_D.$

The process of recollection: starting with the probability
distribution $\mu$  the brain computes the covariance operator $D$
of  this probability distribution by using the ensemble averaging,
see (\ref{COV}), (\ref{COV1}) .

On the basis of this covariance operator it produces a signal
$\phi_{\rm{recall}}(s,x), x \in O,$   a trajectory of the
corresponding Gaussian process.

In this situation, the brain does not reproduce the original
signal $\phi(s,x),$ see CSP1.  The graphs of  $\phi(s,x)$ and
$\phi_{\rm{recall}}(s,x)$ can differ essentially point wise.
Moreover, the original signal $\phi(s,x)$ need not be Gaussian at
all. However, correlations inside both signals approximately coincide
 and, hence, their qualia:
$$
\langle f \rangle_\phi = \lim_{T \to \infty} \frac{1}{T} \int_0^T
f(\phi(s)) ds \approx  \lim_{T \to \infty} \frac{1}{T} \int_0^T
f(\phi_{\rm{recall}}(s)) ds= \langle f
\rangle_{\phi_{\rm{recall}}}.
$$

We also remark that even two different recollections
$\phi^{1}_{\rm{recall}}(s,x)$ and $\phi^{2}_{\rm{recall}}(s,x)$ of
the same image can be very different as physical signals -- two
different realizations of the same Gaussian process. However,
their qualia coincide:
\begin{equation}
\label{EOYU} \langle f \rangle_{\mu_{D}}=\lim_{T \to \infty}
\frac{1}{T} \int_0^T f(\phi^{1}_{\rm{recall}}(s)) ds  = \lim_{T
\to \infty} \frac{1}{T} \int_0^T f(\phi^{2}_{\rm{recall}}(s)) ds.
\end{equation}
To be more precise, we say that they coincide approximately, since in reality the brain does not  calculate the limit for $T \to \infty,$
but it uses the finite $T.$ Thus
\begin{equation}
\label{EOYU}
\langle f \rangle_{\mu_{D}} \approx \frac{1}{T} \int_0^T f(\phi^{1}_{\rm{recall}}(s)) ds
\approx  \frac{1}{T} \int_0^T f(\phi^{2}_{\rm{recall}}(s)) ds.
\end{equation}

\medskip

CSP5. {\bf Recognition of images.} Suppose now that some
mental image was saved in the memory: starting with the input
signal $\phi(s,x)$ and through its covariance operator $D(\phi);$
for example, the Moscow Kremlin. I came to Moscow once again and I
look at the Kremlin; this visual input induces a signal $q(s,x).$
Its covariance operator $D(q)$ is produced, see CSP2. It is
compared with covariance operators in the memory to match with the
operator $D(\phi).$ (The model under consideration does not describe
the mechanism of this comparing process; however, see CSP5n.)
Finally, matching of the operators, $D(\phi) \approx D(q),$ is approached. This activates in the memory
the statistical probability distribution $\mu_{D(\phi)}$ in the form of a Gaussian signal
$\phi_{\rm{recall}}(s,x).$ Its makes the feeling of recognition
of the image which was encoded by $D(\phi).$

\subsection{Classical Signal Processing of Mental Images: finite-dimensional approximations}
\label{DFD}

In principle, the brain can calculate the complete covariance
operator (\ref{EOQ!!!}); especially if it works as analogous
computational device. However, it consumes a lot of computational
resources. We might speculate that the brain selects a finite
number of test functions, it is always possible to assume that
they are orthogonal in $L_2(O):$
\begin{equation}
\label{SYST}
u_1(x),...,u_n(x).
\end{equation}
Instead of the complete covariance operator $D= D(\phi),$  the
brain calculates its cutoff, the covariance $n \times n$ matrix
$D_n= D_n(\phi).$ Thus, instead of infinite dimensional
$L_2$-space, the brain works (for a given mental function) in
fixed finite dimensional subspace $H_n.$ We modify CSP1-CSP5; the
first step CPS1 is not changed. We have:

\medskip

CSP2n. The signal $\phi(s,x)$ induces the mental image $MI_{\phi; n}$ encoded by the covariance matrix $D_n.$

CSP3n. The $MI_{\phi; n}$ is recorded in the memory through the
probability distribution $\mu_{D_n}$ on the finite dimensional
Hilbert space $H_n.$

CSP4n. On the basis of $D_n$ the brain produces a signal
$\phi_{\rm{recall}}(t)$ in $H_n.$ Its activation is recollection
of the memory on $MI_{\phi; n}.$

CSP5n. The memory contains the image $MI_{\phi; n}$ in the form of
the matrix $D_n(\phi).$ The new signal produces $MI_{q; n}$ with
the covariance matrix $D_n(q).$ These matrices  must be compared.
Since the whole story is about covariance matrices, so $n\times n$
matrices, it is natural to expect a comparing algorithm which
compares cutoffs of these $n \times n$ covariance matrices: first
of the dimension two, then three and so on; i.e., first the
$D_2(q)$ is compared with  $2\times 2$ matrices obtained through
projection of mental images on $H_2$ until the cluster of matrices
with the left-up block $D_2(q) (=D_2(\phi))$ is found; then inside
this cluster the brain is looking for the sub-cluster with the
left-up block $D_3(q)(=D_3(\phi))$ and so on.

\medskip

How does the brain selects the subspace $H_n$ with the basis (\ref{SYST})?

\medskip

The most natural is to assume that it just selects a band of
frequencies. It is also natural that different mental functions
may use different bands, i.e., different Hilbert spaces:  for
mental functions  $F$ and $G,$ two Hilbert spaces $H_F$ and $H_G.$

\section{Quantum-like processing of mental images}
\label{QLQL}

We still proceed with functioning of one fixed mental function, say $F.$

\medskip

QLP1. {\bf Density operator code.} At some stage of its growing a
cognitive system creates a sufficiently extended database of
classical mental images. They are encoded by covariance operators
which are transferred in density operators by normalization, see
aMI, section \ref{LKP}. Thus the brain created a collection of QL
mental states borrowed from the classical processing, $\rho \in
{\cal D}_{\rm{data}}(H)$ which is a subspace of ${\cal D}(H).$ At
this stage the brain can be fine by working inside ${\cal D}(H),$
i.e., even without contacts with physical and mental environment.

\medskip

QLP2. {\bf Unitary thinking.} Processing of information inside
${\cal D}(H)$ is the process of QL-thinking. Starting with the
operator $\rho_0$ the brain induces the evolution $\rho(s)$ of the
mental QL state. The simplest dynamics corresponds to the process
of thinking in the absence of inputs from  environment (which
includes the body); it is given by the von Neumann equation:
\begin{equation} \label{EOH} i \frac{d \rho(t)}{d t}= [\widehat{{\cal H}}, \rho(t)], \; \rho(0)= \rho_0,
\end{equation}
where $\widehat{ {\cal H}}: H \to H$ is ``mental Hamiltonian''
(given by a self-adjoint operator). It describes functioning  of
the mental function $F$ under consideration, cf. \cite{}.

In the simplest case Hamiltonian $\widehat{{\cal H}}$ is
completely determined by the mental function $F,$ so
$\widehat{{\cal H}} \equiv \widehat{{\cal H}}_F.$ However, even
more complex dynamics seem to be reasonable -- with $\widehat{
{\cal H}}$ which also depends on the initial state $\rho_0:
\widehat{ {\cal H}}\equiv \widehat{{\cal H}}_{F,\rho_0},$ cf. with
the QL model of decision making, Chapter ?.

We remark that by starting with e.g. QL-version of a concrete image, i.e.,
$\rho_0 \in  {\cal D}_{\rm{data}}(H),$ the QL dynamics can go away from this subspace
of ${\cal D}(H).$  New ``really QL'' images are created. They can be visualized through production of corresponding classical signals, see see bMI, section \ref{LKP}.

We emphasize that the QL-dynamics of mental images is performed on the $T$-scale which is rough comparing with the $\tau$-scale
of physical processing of signals in the brain. Each instant of time $t$ of the $T$-scale is the (large) interval $T$
of the $s$-time.

It may be more illustrative to consider the discrete dynamics, the mental time $t$
is considered as the discrete parameter: $t_n= nT.$ Then
\begin{equation}
\label{EOHD} i \rho(t_{n+1})=  T [\widehat{{\cal H}}, \rho(t_n)],
\; \rho(0)= \rho_0.
\end{equation}

QLP3. {\bf Dynamics of QL qualia.} In QL-processing qualia are
reduced to {\it quadratic functionals} of premental (physical)
signals. These functionals are represented by their
QL-counterparts -- corresponding self-adjoint operators. The
evolution of QL qualia is described by the Heisenberg equation:
\begin{equation}
\label{EOHH} - i \frac{d \widehat{A}(t)}{d t}= [\widehat{{\cal
H}}, \widehat{A}], \; \widehat{A}(0)= \widehat{A}_0;
\end{equation}
or in the discrete representation of the mental time:
\begin{equation}
\label{EOHD} i \widehat{A}(t_n)=  T [\widehat{{\cal H}},
\widehat{A}(t_n)], \; \widehat{A}(0)= \widehat{A}_0.
\end{equation}
Quale (encoded by $\widehat{A})$  of a mental image $MI_\rho$ (encoded by the density operator
$\rho)$ is quantified by its average given by the quantum formula (\ref{AV1}).

Of course, the transition from the class of  classical qualia
(given by arbitrary functionals of signals) to QL mental features
corresponding to only quadratic functionals simplifies mental
representation of an image. However, this reduction can be
justified by (\ref{AV3Z}) in the framework of QLR  (\ref{AV3Z0})
of classical functionals of signals.

\medskip

QLP4. {\bf Thinking via operator algebra.} This is the crucial
point. In QL-thinking the brain switches from the classical
physical signal processing, i.e., nonlinear equations of the type
(\ref{PRSIG}) to {\it linear processing} of mental images
represented by density operators (\ref{EOHD}); the representation
of qualia is also essentially simplified and it can be done in the
linear operator form. Our conjecture is that the brain is really
able to realize such linear operator processing of mental
entities. This type of processing is especially profitable for
``abstract thinking'', i.e., thinking which has a high degree of
independence from inputs.

How does the brain realize the QL (operator) processing on the physical level?

We do not know yet. However, we hope that our model may stimulate
neurophysiologists to look for the corresponding neuronal
representation of QL-processing. We can present the following
scheme of mental operator processing:

Since the brain has no other computational resources different
from neural electric activity, it seems reasonable to assume that
the QL mental dynamics (\ref{EOH}), (\ref{EOHD}) also has to be
performed through this activity. The production of density
operators can be done similarly to the production of covariance
operators in the classical regime. The only difference is that the
brain wants to escape the complicated nonlinear evolution
(\ref{PRSIG}). We consider the following stochastic linear
dynamics in Hilbert space $H$ (of classical electromagnetic
fields):
\begin{equation}
\label{SCHSCH} \frac{\partial \phi}{\partial t}(t,x,\omega) =
\widehat{{\cal H}}\phi (t,x,\omega), \;\phi (t_0,x,\omega)= \phi_0
(x,\omega),
\end{equation}
where the random variable $\phi_0 (x,\omega)$ is the Gaussian
field with zero mean value and the covariance operator $\rho_0.$
Hence $\langle \rho_0 u, v\rangle = E\langle \phi_0, u \rangle
\langle v,  \phi_0\rangle, u,v \in H.$ The solution of the Cauchy
problem (\ref{SCHSCH}) is the random field:
\begin{equation}
\label{SCHSCH1} \phi(t,x,\omega)= U_t \phi_0(x,\omega),
\end{equation}
where $u_t=e^{-it \widehat{{\cal H}}}$ is the standard for QM one
parametric group of unitary operators. The covariance operator
$\rho(t)\equiv \rho_{\phi(t)}$ can be easily found: $\langle
\rho(t) u, v\rangle = E \langle U_t \phi_0, u \rangle \langle v,
U_t \phi_0\rangle =  E \langle \phi_0, U_t^*u \rangle \langle
U_t^* v, \phi_0\rangle = \langle \rho(t) U_t^* u, U_t^*v\rangle.$
Thus
$$
\rho(t)= U_t \rho_0 U_t^*.
$$
This operator-valued function  $\rho(t)$ satisfies the von Neumann
equation (\ref{EOH}). Thus the von Neumann evolution of the mental
state can be induced by the linear dynamics with random initial
condition (\ref{SCHSCH}). As was mentioned, the crucial point is
that this dynamics is much simpler than the  ``classical signal
dynamics'' (\ref{PRSIG}).

Finally, we have the following model of physical realization of
the evolution (\ref{EOH}). In fact, the brain  produces the
Gaussian random signal by realizing on the neuronal level the
linear Schr\"odinger type evolution. At each moment of mental time
$t$ by calculating its covariance operator the brain creates the
mental image given by the covariance-density operator $\rho(t).$

In quantum information theory it is well known that in general ,
i.e., in the presence of interaction with environment the von
Neumann equation should be modified to the
Gorini-Kossakowski-Sudarshan-Lindblad equation:
\begin{equation}
\label{EOH} i \frac{d \rho(t)}{d t}= \widehat{L} (\rho(t)), \;
\rho(0)= \rho_0,
\end{equation}
where $\widehat{L}: {\cal L}(H) \to {\cal L}(H)$ is a linear map
with special properties, see \cite{}. All previous considerations
can be easily generalized to such mental dynamics.

\medskip

QLP5. {\bf Concepts.} Consider a subspace $L$ of $H$ and the orthogonal projector
$\pi\equiv \pi_{HL}: H \to L.$ It induces the map  $\pi: {\cal D}(H) \to {\cal D}(L),$
$$
\rho_L\equiv \pi(\rho)= \frac{\pi \rho \pi}{\rm{Tr} \pi \rho \pi}.
$$
Mental images corresponding to elements of ${\cal D}(L)$ can be considered as abstractions of
mental images corresponding to elements of ${\cal D}(H);$ we call them $L$-concepts or simply concepts.
 Take some $\rho_L\in  {\cal D}(L).$
The mental image $MI_{\rho_L}$ can be interpreted as an abstract
concept induced by the cluster of mental images:
$$
W_{\rho_L}= \{ \rho \in {\cal D}(H): \pi(\rho) = \rho_L\}.
$$
Each concept is based on common correlations of a cluster of
mental images. It is especially interesting to consider the case
$\rm{dim}\; L=m$ and $m$ is quite small. These are very abstract
concepts which contains only the basic common correlations in a
huge cluster of mental images. It is extremely profitable for the
brain to think on the conceptual level; especially to operate in a
finite-dimensional $L.$ The operator unitary dynamics (\ref{EOH})
is reduced to the matrix dynamics. Conceptual Hamiltonian is given
by a symmetric $m \times m$ matrix. For small $m,$ dynamics of
such a type are very simple; processing is very rapid.

For example, consider QLR for the concept ``house''. In the classical regime the brain created a collection of
images of concrete houses $MI_1,..., MI_k.$ They were classically encoded by covariance operators $D_1,..., D_k.$ These operators contain some common
correlations. In the matrix representation they have a common block. For simplicity, suppose that this block is of the diagonal type. Consider a subspace $L$ of $H$ related to this block. Then this block can be represented as a self-adjoint
and positive operator  $D_L$ in $L.$ We have:
$$
D_L= \pi D_j \pi, j=1,2,..., k,
$$
where $\pi: H \to L$ is the orthogonal projector. Its QL image is given by
$$
Q_L= D_L/\rm{Tr} D_L= \frac{\pi D_j \pi}{ \rm{Tr} \pi D_j \pi}=
\frac{\pi (D_j/ \rm{Tr} D_j)\pi}{ \rm{Tr} \pi (D_j/ \rm{Tr} D_j) \pi}=
\frac{\pi \rho_j\pi}{ \rm{Tr} \pi \rho_j \pi} = \rho_L.
$$
Here $\rho_j$ are QL representations of the covariance operators $D_j.$
The density operator $\rho_L$ gives the abstract concept of house.

We now  describe the process of creation of a concept of a higher level of abstractness
from a cluster of concepts. Consider a subspace $Z$ of $L.$ The brain can create new concepts
belonging to ${\cal D}(Z)$ starting with clusters of $L$-concepts. Of course, it can proceed directly
starting with mental images from  ${\cal D}(H).$ However, such step by step increasing of the level of
abstraction is very natural.

\medskip

QLP6. {\bf Neuronal location of the QL-processor.} From the
general viewpoint there are no reasons to assume that QLR is
realized in the same physical domain, the same ensemble of
neurons, as classical processing. It may be that there is a
special domain $O_{QL}$ which is used for dynamics (\ref{EOH}).
Our model induced an interesting problem of experimental
neurophysiology -- to find domains of the brain coupled to QLR. If
the hypothesis that the dynamics (\ref{EOH}) of mental images is
based on the physical dynamics (\ref{SCHSCH}) is correct, then
domains of QLR can be identified by the presence of Gaussian
stochastic dynamics. Unfortunately, at the present level of
measurements it is impossible to measure directly the
electromagnetic field inside the brain (at least to make
measurements in a sufficiently dense set of points). However, even
the  measurement technology based on EEG provides a possibility of
reconstruction of the field inside the brain by using the methods
of the inverse problem, \cite{}.

QLP7. {\bf Quantum-like consciousness.}
We may speculate that consciousness can be associated with QL processing in the brain.
The von Neumann equation (or more generally the
Gorini-Kossakowski-Sudarshan-Lindblad equation) represents the ``continuous flow of consciousness''.
The feeling of continuity is generated through averaging of physical signals with respect to the mental time scale.
i.e., the representation of mental images by covariance-density operators. In fact, on the physical time scale the
dynamics is discrete, see (\ref{EOHD}).

QLP7. {\bf Correspondence between classical and quantum qualia.}  Consider a classical quale given by a
functional $f(\phi).$ The corresponding quantum quale is given by the self-adjoint operator, the second derivative
of $f(\phi)$ at the point $\phi=0.$ In general behavior of the functional $f(\phi)$ differs essentially from
behavior of its quadratic part. However, on the level of averages the difference is not so large: in the limit
$\tau/T \to 0$ they coincide, see (\ref{AV3Z}).

\section{Composite  systems}
\label{CS}

We turn again to physics. In CSM a composite system $S=(S_1, S_2)$ is mathematically
described by the Cartesian product of state spaces of its parts
$S_1$ and $S_2.$ In QM it is described by the tensor product.
Majority of researchers working in quantum foundations and,
especially quantum information theory, consider this difference in
the mathematical representation as crucial. In  particular,
entanglement which is a consequence of the tensor space
representation is considered as totally nonclassical phenomenon.
However, we recall that Einstein considered the EPR-states as
exhibitions of  classical correlations due to the common
preparation. PCSFT will realize Einstein's dream on entanglement.

Let $S=(S_1, S_2),$ where $S_i$ has the state space $H_i$ --
complex Hilbert space. Then by CSM the state space of $S$ is
$
H_1\times H_2.
$
By extending PCSFT to composite systems we should describe
ensembles of composite systems by probability distributions on
this Cartesian product, or by a random field $\phi(x,\omega)=
(\phi_1(x,\omega), \phi_1(x,\omega)) \in H_1\times H_2.$

In our approach each quantum system is described by its own random field: $S_i$ by
$\phi_i(x,\omega), i=1,2.$ However, these fields are CORRELATED -- in completely classical sense.
Correlation at the initial instant of time $s=s_0$ propagates in time in the complete accordance with
laws of QM. There is no action at the distance. It is a purely classical dynamics of two stochastic processes which were
correlated at the beginning. (In fact, the situation is more complex: there is also the common random background,
vacuum fluctuations; we shall come back to this question a little bit later).

\subsection{Operator realization of wave function}

Consider now the QM-model, take a pure state case: $\Psi \in H_1\otimes H_2.$
Can one peacefully connect the QM and PCSFT formalisms? Yes! But
$\Psi$ should be interpreted in completely different way than in the conventional QM.

The main mathematical point: $\Psi$ is not
vector! It is an operator! It is, in fact, the non-diagonal block of
the covariance operator of the corresponding prequantum random
field: $\phi(x,\omega) \in H_1 \times H_2.$
The wave function $\Psi(x,y)$ of a composite system determines the integral operator:
$$
\widehat{\Psi} \phi(x)= \int \Psi(x,y) \phi(y) dy.
$$
We keep now to the finite-dimensional case.
Any vector $\Psi \in H_1\otimes H_2$  can be represented in the
form $\Psi= \sum_{j=1}^m \psi_j \otimes \chi_j, \; \psi_j \in H_1,
\chi_j \in H_2,$
and it determines a linear operator from $H_2$ to $H_1$
\begin{equation}\label{TP1}
\widehat{\Psi} \phi= \sum_{j=1}^m (\phi, \chi_j) \psi_j, \; \phi
\in H_2.
\end{equation}
 Its
adjoint operator $\Psi^*$ acts from $H_1$ to $H_2:$
$\widehat{\Psi}^* \psi= \sum_{j=1}^m (\psi, \psi_j) \chi_j , \psi
\in H_1.
$
Of course, $\widehat{\Psi} \widehat{\Psi}^*: H_1 \to H_1$ and
$\widehat{\Psi}^* \widehat{\Psi}: H_2 \to H_2$ and these operators
are self-adjoint and positively defined.
Consider  the density operator corresponding to a pure quantum state, $\rho= \Psi \otimes \Psi.$ Then the operators of the partial traces
$\rho^{(1)}\equiv \rm{Tr}_{H_2} \rho= \widehat{\Psi}
\widehat{\Psi}^*$ and $\rho^{(2)}\equiv \rm{Tr}_{H_1} \rho
=\widehat{\Psi}^* \widehat{\Psi}.$

\subsection{Basic equality}

Let $\Psi\in \in H_1\otimes H_2$  be normalized by 1. Then, for
any pair of linear bounded operators $\widehat{A}_j: H_j \to H_j , j= 1,2,
$ we have:
\begin{equation}
\label{01} \rm{Tr} \widehat{\Psi} \widehat{A}_2 \widehat{\Psi}^*
\widehat{A}_1= \langle \widehat{A}_1 \otimes \widehat{A}_2
\rangle_\Psi \equiv (\widehat{A}_1 \otimes \widehat{A}_2 \Psi,
\Psi).
\end{equation}
This is a mathematical theorem\cite{PCSFT4}; it will play a fundamental role in further considerations.

\subsection{Coupling of classical and quantum
correlations} \label{PSI}

 In PCSFT a composite system $S=(S_1,
S_2)$ is mathematically represented by the random field $\phi
(\omega) = (\phi_1 (\omega), \phi_2 (\omega)) \in H_1\times H_2.$
Its covariance operator $D$
 has the block structure
\[D = \left( \begin{array}{ll}
 D_{11} & D_{12}\\
D_{21} & D_{22}\\
 \end{array}
 \right ),
 \]
 where $D_{ii}: H_i \to H_i, D_{ij}: H_j \to H_i.$
The covariance operator is self-adjoint. Hence $D_{ii}^* = D_{ii},$ and
$D_{12}^* = D_{21}.$

Here by the definition:
 $
(D_{ij} u_j, v_i) = E (u_j, \phi_j (\omega)) (v_i, \phi_i
(\omega)), u_i \in H_i, v_j \in H_j.
$
For any Gaussian random
vector  $\phi (\omega) = (\phi_1 (\omega), \phi_2 (\omega))$
having zero average
 and any pair of operators
$\widehat{A}_i \in {\cal L}_s (H_i), i= 1,2,$ the following
equality takes place:
 $
   \langle f_{A_1}, f_{A_2}\rangle_\phi \equiv
 E f_{A_1}(\phi_1 (\omega)) f_{A_2} (\phi_2 (\omega))
 = ({\rm Tr} D_{11} \widehat{A}_1) ({\rm Tr} D_{22} \widehat{A}_2)
+  {\rm Tr} D_{12} \widehat{A}_2 D_{21} \widehat{A}_1.
$
We remark that
${\rm Tr} D_{ii} \widehat{A}_i = E f_{A_i} (\phi_i (\omega)),
i=1,2.
$
Thus we have
 $ f_{A_1} f_{A_2} =  E f_{A_1} E f_{A_2} +
{\rm Tr} D_{12} \widehat{A}_2 D_{21} \widehat{A}_1.
$
Consider a Gaussian  vector random field such
that $ D_{12}= \widehat{\Psi}:$
 \begin{equation}
 \label{00t5}
E (f_{A_1} - E f_{A_1}) (f_{A_2} - E f_{A_2}) = (\widehat{A}_1
\otimes \widehat{A}_2 \Psi, \Psi) \equiv \langle \widehat{A}_1
\otimes \widehat{A}_2 \rangle_\Psi,
\end{equation}
or, for covariance of two classical random vectors $f_{A_1},
f_{A_2},$ we have:
${\rm cov} \; (f_{A_1}, f_{A_2}) = \langle
\widehat{A}_1 \otimes \widehat{A}_2 \rangle_\Psi.
$

We have the following equality for averages of quadratic forms of coordinates of the prequantum random field describing the
state of a composite system:
$E f_{A_i} (\phi_i) (\omega)) = {\rm Tr} D_{ii} \widehat{A}_i.$
We want to construct a random field such that these averages will match those given by
QM. For the latter, we have:
$
\langle \widehat{A}_1 \rangle_\Psi = (\widehat{A}_1 \otimes I_2
\Psi, \Psi) = {\rm Tr} (\Psi \Psi^*) \widehat{A}_1; $$ $$ \langle
\widehat{A}_2 \rangle_\Psi =
 (I_1 \otimes \widehat{A}_2 \Psi, \Psi) = {\rm Tr} (\widehat{\Psi}^* \widehat{\Psi})
\widehat{A}_2,
$
where $I_i$ denotes the unit operator in $H_i, i=1,2.$ Thus it
would be natural to take
$D_\Psi= \left( \begin{array}{ll}
 \widehat{\Psi} \widehat{\Psi}^* & \; \; \widehat{\Psi} \\
 \; \; \widehat{\Psi}^* & \widehat{\Psi}^*\Psi
 \end{array}
 \right ).
 $
However, {\it this operator is not positively defined!}
It could not determine any probability distribution on the space of classical fields. We modify it to obtain a positively defined operator. Originally this modification had purely mathematical reasons, but there are deep physical grounds for it.

The operator
$\tilde{D}_\Psi= \left( \begin{array}{ll}
 \widehat{\Psi} \widehat{\Psi}^* + \epsilon I & \; \; \; \; \; \; \; \widehat{\Psi}\\
\; \;  \; \; \; \; \; \widehat{\Psi}^* & \widehat{\Psi}^*\Psi +
\epsilon I
 \end{array}
 \right )
 $
is positively defined if $\epsilon > 0$ is large enough \cite{PCSFT4}.
 Hence, it determines uniquely the Gaussian measure on the space of classical fields.
Suppose now that $\phi (\omega)$ is a random vector with the
covariance operator $\tilde{D}_\Psi.$ Then
\begin{equation}
\label{YY1} \langle \widehat{A}_1 \rangle_\Psi = E f_{A_1} (\phi_1
(\omega)) - \epsilon {\rm Tr} \widehat{A}_1.
 \end{equation}
This relation for averages and relation (\ref{00t5})
provide coupling between PCSFT and QM. Quantum statistical
quantities can be obtained from corresponding quantities for
classical random field:
{\it ``irreducible quantum randomness" is
reduced to randomness of classical prequantum fields.}

\subsection{Vacuum fluctuations}

The additional term given by the unit operator in the diagonal blocks of the covariance operator of the
prequantum vector field corresponds to the field of the white noise type. Such a field can be considered as
vacuum fluctuations, vacuum field.
 PCSFT induces the following picture of reality:

Fluctuations of the vacuum field  are combined with random fields
 representing quantum systems.  Since we cannot separate, e.g., electron from the vacuum field,
we cannot separate totally any two quantum systems. Thus all quantum systems
are ``entangled'' via the vacuum field.

\subsection{Superstrong quantum correlations}
\label{STR}

In PCSFT such correlations (violating Bell's inequality) are due to the presence of the vacuum field.  The off-diagonal
term $\widehat{\Psi}$ can be so large only if the diagonal terms are completed by the contribution of the vacuum filed.
Mathematics tells us this. Thus they are so strong, because the vacuum field really couple any two systems;
they are in the same fluctuating space.

Space is a huge random wave; quantum systems are spikes on this wave; they are correlated
via this space-wave. Thus quantum correlations have two contributions:

\medskip

1) initial preparation;

2) coupling via the vacuum field.

\medskip

The picture is pure classical... In this model the vacuum field is the source of additional correlations.
It seems that this classical vacuum field is
an additional (purely classical) quantum computational resource.

\section{Entanglement of mental functions}

Consider now two mental functions $F_1$ and $F_2$ which are spatially coupled to domains $O_1$ and $O_2$
of the brain.

\subsection{Classical correlated processing of mental images}

Consider the input signal  $\phi_0$ which is split into
sub-signals used by $F_1$ and $F_2,$ respectively: $\phi_0=
(\phi_{01}, \phi_{02}).$ We emphasize that these are correlated
signals; i.e., in general the nondiagonal term of their covariance
matrix $D_0$ is nonzero. Starting with $\phi_0$ the $F_1$ and
$F_2$ produce the random field $\phi(t,x, \omega) = (\phi_1(t,x,
\omega), \phi_2(t,x, \omega)).$ In the general case  $F_1$ and
$F_2$ can interact on the physical level. But in principle they
can proceed their signals $\phi_1(t,x, \omega)$ and  $\phi_2(t,x,
\omega)$ without any physical interaction (with negligibly small
interaction). For example, domains $O_1$ and $O_2$  are
sufficiently separated in space. We shall concentrate our study on
such a case, since here the QL features of the model are the most
visible.

Thus the brain produces the covariance operator $D(t)=
D_{\phi(t)}$ representing the classical mental image at the moment
$t$ of mental time. Its dynamics is very complicated, it is
induced by in general nonlinear dynamics of corresponding signals
$\phi_1(t,x, \omega)$ and $\phi_2(t,x, \omega).$

We now coming to an important issue of our model, namely, the {\it
role of noise in information processing.} The presence of noise in
the brain can be considered as a disturbing effect for
deterministic information processing. However, in our model mental
images are encoded by covariance operators. Incorporation of the
noisy contribution in the covariance operator does not induce
problems. Moreover, it can improve (!) information processing by
coupling through noise  spatially separated processes in the
brain.

Let us consider noise of the white noise type, i.e., the Gaussian
random field with zero mean value and the covariance operator
$D_{\rm{noise}} = \epsilon I, \epsilon >0.$ We now suppose that
this noise is incorporated into the signal $\phi(t,x, \omega)$
produced by  $(F_1,F_2).$  Hence,
$$
D(t)= \tilde{D}(t) + D_{\rm{noise}}.
$$
We remark that nondiagonal blocks describing correlations between
mental functions $F_1$ and $F_2$  do not depend on the presence of
the noise. However, the presence of the noise provides a
possibility to make these nondiagonal blocks essentially larger
(in the operator sense) than in its absence. Thus the noise
coordinates processes in $O_1$ and $O_2:$ it produces so to say
nonlocal correlations.

\subsection{Quantum-like entangled processing of mental images}
\label{ENTPR}
We now can repeat the QL-story on the information processing by
considering mental images encoded by density operators of the
class ${\cal D}(H_1 \otimes H_2).$ It is possible to embed this
space into the space of all covariance operators on the Cartesian
product $H_1 \times H_2.$ We have done this for the density
operators corresponding to pure quantum states $\Psi \in H_1
\otimes H_2,$ see section \ref{PSI}. It is possible to generalize
our construction to arbitrary density operators in $H_1 \otimes
H_2,$ see \cite{}. The dynamics of the mental state is described
by von Neumann equation (or more generally the
Gorini-Kossakowski-Sudarshan-Lindblad equation) in ${\cal D}(H_1
\otimes H_2).$  Each pair of qualia $\widehat{A}_1$ and
$\widehat{A}_2$ associated with the mental functions $F_1$ and
$F_2$ forms a new quale which is given by the operator
$\widehat{A}_1\otimes \widehat{A}_2.$ It is quantified by its
average.

The neuronal realization of the QL-dynamics can be performed in
the same way as it was done for a single mental function, see
QLP4. Consider the case of two isolated processing: in $O_1$ and
$O_2.$ Each of them is described by mental Hamiltonian
$\widehat{{\cal H}}_i: H \to H, i=1,2.$ In this case QM  dynamics
is given by the Schr\"odinger equation with Hamiltonian
$\widehat{{\cal H}}= \widehat{{\cal H}}_1 \otimes I + I \otimes
\widehat{{\cal H}}_2:$
\begin{equation}
\label{SCHSCHj} \frac{\partial \psi}{\partial t}(t,x_1,x_2) =
\widehat{{\cal H}}\psi (t,x_1,x_2, \;\psi (t_0,x_1,x_2)= \psi_0
(x_1,x_2).
\end{equation}
We remark that, since in general the initial wave function is not
factorizable, i.e., $\psi (x_1,x_2) \not= \psi_{01} (x_1)
\psi_{02} (x_2),$ the Schr\"odinger equation cannot be split into
a system of equations corresponding to the subsystems of the
composite system. However, we can do this on the level of the
prequantum stochastic process and obtain a system of linear
equations with random initial conditions, $j=1,2:$
\begin{equation}
\label{SCHSCHj1} \frac{\partial \phi}{\partial t}(t,x_j, \omega) =
\widehat{{\cal H}}_j \phi (t,x_j, \omega), \;\phi(t_0,x_j,
\omega)= \psi_0 (x_j, \omega).
\end{equation}
The covariance operator $\rho(t)=\rho_{(\phi_1(t), \phi_2(t))}$
induced by the solution of this system is the solution of the von
Neumann equation for the composite system.

Correlations between processes related to $F_1$ and $F_2$ are
suprestrong! They can violate Bell's inequality! However, in our
model this is not a consequence of mystical quantum nonlocality,
but of the presence of the white noise which increases the
correlation effect.

In ``prequantum physics''  this noise corresponds to the
background (zero point field), it represents vacuum fluctuations.
In the brain this is the ordinary macroscopic noise of a huge
electric network consisting of billions of neurons.

\end{document}